\documentclass[review]{elsarticle}

\usepackage{lineno,hyperref}
\modulolinenumbers[5]

\journal{Journal of \LaTeX\ Templates}

\begin{document}
\title{
 Bicriteria Optimization  in  Designing Magnetic Composites
}

\author{Krzysztof Z. Sokalski}\ead{ksokalski76@gmail.com}
\address{Institute of Computer Science, Cz\c{e}stochowa University of Technology,
Al. Armii Krajowej 17, 42-200 Cz\c{e}stochowa, Poland}

\author{Barbara \'{S}lusarek}\ead{barbara.slusarek@itr.org.pl},\author{Bartosz Jankowski}\ead{bartosz.jankowski@itr.org.pl },\author{Marek Przybylski}\ead{marek.przybylski@itr.org.pl }

\address{Tele and Radio Research Institute, Ratuszowa 11 Street, 03-450 Warszawa, Poland} 



\date{}
\begin{abstract}
 A novel algorithm for designing values of technological parameters for production of soft magnetic composites (SMC) has been created. These parameters are hardening temperature $T$ and compaction pressure $p$. They enable us to optimize power losses and induction. The advantage of the presented algorithm consists in bicriteria optimization. The crucial role played by the presented algorithm is the scaling and pseudo-state equation. On this basis mathematical models of power losses and induction have been created. The model parameters have been calculated on the basis of the power loss characteristics and hysteresis loops. The created optimization system has been applied to specimens of Somaloy 500. The obtained output consists of a finite set of feasible solutions. In order to select a unique solution an additional criterion has been formulated.
\end{abstract}

\maketitle

\section{Introduction}\label{I}
Soft magnetic composites (SMC) have physical properties which are used for adapting these materials to specific applications \citep{bib:Shok},\citep{bib:Lemi}. Very often the functionality of these materials depends on more than one feature. This leads to multi-criteria optimization problems,   which has not been applied yet in the design of SMC. However, there are papers which treat more than one physical property of SMC but these are not considered as target functions in an optimization procedure \citep{bib:Sho2},\citep{bib:Sho3}. 
Recently an algorithm for designing values of the hardening temperature and the compaction pressure in the production process of soft magnetic composites (SMC) has been derived by using the concept of the pseudo-state equation \citep{bib:sok2}. In equilibrium thermodynamics the equation of state relates thermodynamic parameters. For instance, in the case of gas-liquid system they are the temperature, pressure and volume of the considered material. By analogy with the equation of state we consider a phenomenological relation between the technological parameters and physical properties of the material. Such an approach for SMC is possible thanks to the topology of the completed set of scaled power loss characteristics. The most important features of this topology are the following,\citep{bib:sok2}:

Set of characteristics consists of one variable smooth function
\begin{equation}
\label{uno}
\frac{P_{tot}}{(B_{m})^{\beta}}=F\left (\frac{f}{B_{m}^{\alpha}}\right ),
\end{equation}
where $P_{tot}$ is density of power loss, $B_{m}$ is peak of induction, $f$ is frequency of electromagnetic field wave, $F(\cdot)$ is a function of the following form \citep{bib:sok2}:
\begin{equation}
\label{equ}
F\left (\frac{f}{B_{m}^{\alpha}}\right )=(f/B_{m}^{\alpha}\cdot\left(\Gamma_{1}+f/B_{m}^{\alpha}\cdot\left(\Gamma_{2}+f/B_{m}^{\alpha}\cdot\left(\Gamma_{3}+
f/B_{m}^{\alpha}\cdot\Gamma_{4}\right)\right)\right),
\end{equation}
where $\Gamma_{i}, \alpha$ and $\beta$ have to be determined from experimental data. The form (\ref{uno}) has been derived from the assumption about the power losses as a homogeneous function in a general sense.
Each characteristic is determined by the values of $\Gamma_{i}$ coefficients and $\alpha$ as well as $\beta$ exponents. These are functions of the technological parameters $T$ and $p$
\begin{equation}
\label{expony1}
\Gamma_{i}=\Gamma_{i}( T,p),\hspace{2mm}\alpha=\alpha( T,p),\hspace{2mm} \beta=\beta(T,p),
\end{equation}
where $T$ and $p$ are hardening temperature and compaction pressure, respectively. (\ref{expony1})  reveals that characteristics (\ref{uno}), (\ref{equ}) of samples composed at different $T,p$ conditions possess different dimensions, whereas all disentangled characteristics
\begin{equation}
\label{equ2}
P_{tot}=B_{m}^{\beta}(f/B_{m}^{\alpha}\cdot\left(\Gamma_{1}+f/B_{m}^{\alpha}\cdot\left(\Gamma_{2}+f/B_{m}^{\alpha}\cdot\left(\Gamma_{3}+
f/B_{m}^{\alpha}\cdot\Gamma_{4}\right)\right)\right),
\end{equation}
possess a common physical dimension.
 Why do we use the implicit form (\ref{uno}),(\ref{equ})? Note that the right-hand side of this equation depends only on one effective variable $\frac{f}{B_{m}^{\alpha}}$. Therefore, calculations performed with (\ref{uno}),(\ref{equ}) are represented by one curve for all values of $f$ and $B_{m}$, whereas results of calculations performed with (\ref{equ2}) are split into many curves. For instance, if one needs $P_{tot}$ as a function of $f$ then the number of generated characteristics is equal to the required number of different values of $B_{m}$. 
 Because of the different dimensions of the different characteristics they do not cross each other except at origin point $\frac{f}{B_{m}^{\alpha}}=0$, for which dimension is not very important. According to the Egenhofer theorem \citep{bib:egenh} the relations between characteristics are invariant with respect to scaling, translation and rotation. Just the conservation of the relations with respect to the scaling enables us to use the implicit form of characteristics.
All the characteristics are monotonic increasing functions of $\frac{f}{B_{m}^{\alpha}}$
According to (\ref{expony1}) the power loss characteristics are parameterized by pressure and temperature. This dependence enables us to introduce a measure of distance in the space of characteristics. Let $(p_{1},T_{1})$ and $(p_{2},T_{2})$ be labels of the characteristics of the two composites which have been composed under conditions corresponding to these pressures and temperatures, respectively. Then the distance between these characteristics has the following general form:
\begin{equation}
\label{distance}
\rho(p_{1},T_{1},p_{2},T_{2})=R(|p_{1}-p_{2}|,|T_{1}-T_{2}|),
\end{equation}
where $R(\cdot,\cdot)$ satisfies axioms of the distance function. 
Therefore, the set of all characteristics constitutes metric space.We have shown in \citep{bib:sok2} that this space consists of two subspaces. Therefore, by introducing the distance measure in the space of all characteristics we make each of these subspaces compact. 

\begin{figure}
\begin{center}
\includegraphics[width=10cm]{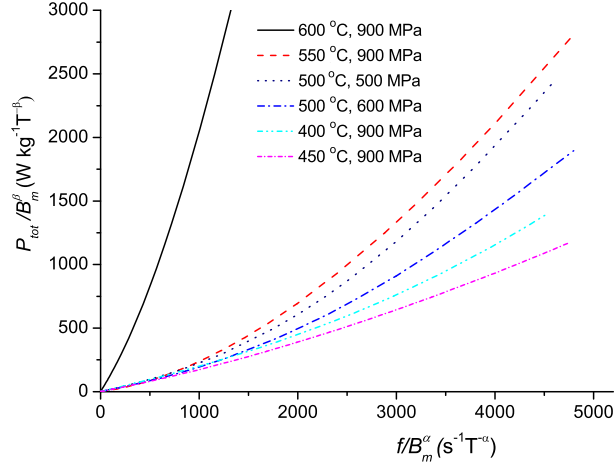}

\caption{The five loss characteristics for $T\le 550^{o}C$ corresponding to the low losses phase and the one characteristic for $T=600^{o}C$ corresponding to the high losses phase.}
\label{H_L_phas}
\end{center}
\end{figure}

Each compact set corresponds to  physical phase which is defined by characteristic values of the physical parameters. For instance, in \citep{bib:sok2} we considered the low and high losses phases of SOMALOY 500.   
All the properties mentioned above are presented in Fig. \ref{H_L_phas}; however, the compactness of characteristics' subsets is ensured by the existence of (\ref{distance}).
These properties have enabled us to introduce a measure of power loss $V(T,p)$ which was the average of characteristics with respect to $\frac{f}{B_{m}^{\alpha}}$ \citep{bib:sok2}:
\begin{equation}
\label{V1} 
V( {T}, {p})=\frac{1}{\phi_{max}-\phi_{min}}\int_{\phi_{min}}^{\phi_{max}}\frac{P_{tot}(\frac{f}{B_{m}^{\alpha}})}{B_{m}^{\beta}}\,d(\frac{f}{B_{m}^{\alpha}}).
\end{equation}
Note that the dimension of the denominator in front of the integral and the dimension of integration limits cancel themselves out.

From the topological properties of the characteristics' set and on (\ref{expony1}) as well as on (\ref{V1}) the pseudo-state equation for soft magnetic composites has been derived. This equation has enabled us to determine the optimal values of the technological parameters \citep{bib:sok2}.
However, the described optimization relates only to the power losses and in design processes optimization of induction is also important. Therefore, the goal of this paper is to create a new algorithm for optimization of both the power losses and the induction within the frame of a bicriteria problem. 
\section{Experimental Data}
Specimens were produced by cold pressing under pressure of 500$\dots$900 MPa. The specimens made of Somaloy 500 powder were cured at a temperature of 400$\dots$600$^o$C for 30 minutes in air atmosphere.
The specimens used in experiments were ring-shaped with a square cross-section. The specimens had the following dimensions: external diameter 55 mm, internal diameter 45 mm and thickness 5 mm.
Total power loss density $P_{tot}$, expressed in watts per kilogram (W/ kg), was obtained from measurements of the AC hysteresis cycle according to IEC Standard 60404-6 using the system AMH-20K-HS produced by Laboratorio Elettrofisico Walker LDJ Scientific. Total power losses$ P_{tot}$ were measured at maximum flux density $B_{m}$ = 0.1\dots1.3 T over a frequency range of 10 to 5000 Hz. During measurements of the total power losses $P_{tot}$, the shape factor of the secondary voltage was equal to 1.111$\pm$1.5 \%. Maximum measurement error of the total energy losses was equal to 3\%.
In order to optimize the magnetic properties, the magnetic inductions $B$ at fixed magnetic field $H$ equal to 1000 A/m were determined. These values were obtained from measurements of the DC magnetization curve according to IEC Standard 60404-4 using the same measuring system. 

\section{Power Losses and Induction Pseudo-State Equations}
Optimization of the power losses was based on the topological properties of the characteristics. However, in the case of magnetic properties the situation is much simpler. For optimization of magnetic properties we selected induction $B_{1000}$ for the fixed magnetic field $H$=1000$($A/m$)$. We chose this value because the magnetic permeability of the soft magnetic composites reaches a maximum value around this magnetic field.
We expected that the pseudo-state equation would properly describe induction at this point as a function of $T$ and $p$. In the previous paper \citep{bib:sok2} it was assumed and confirmed that the loss measure $V$ obeys the scaling. Here, this assumption was extended to induction. In order to justify this assumption we referred to two phenomena: invariance of power losses (area of the hysteresis loop) with respect to scaling  and invariance of the hysteresis loop with respect to scaling \citep{bib:sok3}. Therefore, for the bicriteria optimization problem, minimization of the power losses and maximization of the induction for a fixed magnetic field, we 
used the following pseudo-state equations of general form:


\begin{equation}
\label{general2}
V\left(T,p\right)=\left(\frac{p}{p_{c}}\right)^{\gamma}\,\cdot\,\Phi(X),
\end{equation}
\begin{equation}
\label{general3}
B_{1000}\left(T,p\right)=\left(\frac{p}{p'_{c}}\right)^{\gamma'}\,\cdot\,\Lambda(X'),
\end{equation}
where
\begin{eqnarray}
X=\frac{\frac{T}{T_{c}}}{(\frac{p}{p_{c}})^{\delta}}\label{general2p},\label{general2a}\\
X'=\frac{\frac{T}{T'_{c}}}{(\frac{p}{p'_{c}})^{\delta'}}\label{general3p},\label{general3a}
\end{eqnarray}

where $\Phi(\cdot)$ and $\Lambda(\cdot)$ were arbitrary functions to be determined. $\gamma$, $\delta$ , $\gamma'$, $\delta'$ and $T_{c}$, $p_{c}$, $T'_{c}$, $p'_{c}$ are scaling exponents and scaling parameters respectively, and were to be determined. In the case of the power losses' pseudo-state equation all calculations concerning modelling of $\Phi(\cdot)$ and fitting of scaling exponents as well as model parameters were done in \citep{bib:sok2}. The most important result was the derivation of an infinite set of solutions for the technological parameters which minimized the power losses:
\begin{equation}
\label{optim}
\frac{\frac{T}{T_{c}}}{(\frac{p}{p_{c}})^{\delta}}=19,75.
\end{equation}

\section{ Induction Pseudo-State Equation}
In this Section we derive a pseudo-state equation for induction $B_{1000}$ which will constitute a function of the two variables $p$ and $T$. This function and the power losses' pseudo-state equation (\ref{general2}) will enable us to optimize induction and losses together. The optimization criteria are the following: find $V=V_{min}$ and $B_{1000}=B_{1000\,max}$ with respect to $p$ and $T$. 
Deriving in \citep{bib:sok2} the form for $\Phi( \cdot)$ we reveal two phases of Somaloy 500: low losses and high losses. Therefore, in terms of the induction pseudo-state equation we have to take into account this phase separation. Measurement data of $B_{1000}$ vs. $T$ and $p$ are separated into these two phases in Table \ref{Table:Table2}. 

\begin{table}
\begin{center}
\scriptsize
\caption {Somaloy 500. Measure of induction $B_{1000}$  vs. hardening temperature $T$ and compaction pressure  $p$ for magnetic field $H$=1000 $($A/m$)$.}
\label{Table:Table2}
    \begin{tabular}{|c|c|c|}
    \hline
Temperature & Pressure & Induction \\
	\hline
$($K$)$ & $($MPa$)$ & $($T$)$ \\
    \hline
723,15&	800&	0,378\\
773,15&	900&	0,496\\
773,15&	700&	0,483\\
673,15&	800&	0,335\\
773,15&	600&	0,467\\
823,15&	800&	0,546\\
773,15&	500&	0,414\\
\hline
741,15&	764&	0,425\\
773,15&	750&	0,489\\
773,15&	800&	0,504\\
773,15&	650&	0,469\\
773,15&        725& 0,467\\
873,15&        800& 0,568\\   
\hline
    \end{tabular}
\end{center}
\end{table}
The horizontal line between $B_{1000}=0,414($T$)$ and $B_{1000}=0,425($T$)$ indicates the crossover between the low losses phase and the high losses phase. This transition is clearly visible in the jump of the $V(T,p)$ function around the separation line \citep{bib:sok2}. For each phase we assume an independent branch  of the pseudo-state equation in the form of the Pad{\'e} approximant. In order to simplify notations we introduce the following abbreviations:
\begin{equation}
\label{state0}
\pi=\frac{p}{p_{c}},
\hspace{2mm}\pi'=\frac{p}{p'_{c}},
\end{equation}
Expressing $\Lambda(\cdot)$ in (\ref{general3}) by the Pad{\'e} approximant we get the following form for the induction pseudo-equation of state:
 \begin{equation} 
\label {eq8a} 
B_{1000}(T,p)= \pi'^{\gamma'}\frac{\tilde{G}_{0}+\tilde{G}_{1}\,X'+ \tilde{G}_{2}\,X'^2 +
\tilde{G}_{3}\,X'^3 +\tilde{G}_{4}\,X'^4}{1+\tilde{D}_{1}\,X'+ \tilde{D}_{2}\,X'^2 +
\tilde{D}_{3}\,X'^3 +\tilde{D}_{4}\,X'^4}, 
\end{equation}
where $\tilde{G}_{0},\dots,\tilde{G}_{4}, \tilde{D}_{1},\dots,\tilde{D}_{4}$ are parameters of the Pad\'e approximant. All parameters
have to be determined from the data presented in Table \ref{Table:Table2}. The corresponding pseudo-state equation for the power losses has been derived in \citep{bib:sok2}:

 \begin{equation} 
\label {eq8aa} 
V(T,p)= \pi^{\gamma}\frac{G_{0}+G_{1}\,X+ G_{2}\,X^2 +
G_{3}\,X^3 +G_{4}\,X^4}{1+D_{1}\,X+ D_{2}\,X^2 +
D_{3}\,X^3 +D_{4}\,X^4}.
\end{equation}

\section{ Estimation of Parameters for Induction Pseudo-State Equation }
The above-mentioned crossover between low-loss and high-loss phases is observed as 
A sudden change of $V$ between two points:
$[773,15;500,0]$ and $[742,15;764,0]$ Table \ref{Table:Table2}. However, this effect is not seen in the induction magnitude. Therefore in order to have a compact description of the power losses and the induction we take that into account and we divide the data of Table \ref{Table:Table2} into two subsets corresponding to the two respective phases. Minimizations of $\chi^{2}$ for both phases have been performed with Microsoft Excel 2010, where
\begin{equation}
\label{chi2}
\chi^{2}=\sum_{i=1}^{N}\left( B_{1000}(\tau'_{i},\pi'_{i})- {\pi'_{i}}^{\gamma'} \frac{\tilde{G}_{0}+\tilde{G}_{1}\,{X'}_{i}+ \tilde{G}_{2}\,{X'}_{i}^2 +
\tilde{G}_{3}\,{X'}_{i}^3 +\tilde{G}_{4}\,{X'}_{i}^4}{1+\tilde{D}_{1}\,{X'}_{i}+ \tilde{D}_{2}\,{X'}_{i}^2 +
\tilde{D}_{3}\,{X'}_{i}^3 +\tilde{D}_{4}\,{X'}_{i}^4}\right)^{2}, 
\end{equation}
where $N=7$ and $N=6$ for the low-losses and high-losses phases, respectively.
Table \ref{Table:Table3} and Table \ref{Table:Table4} present estimated values of the model parameters for the low-loss and high-loss phases, respectively. 

\begin{table}
\begin{center}
\scriptsize
\caption {Somaloy 500, low-loss phase. Values of the induction pseudo-state equation's parameters and the Pad{\'e} approximant's coefficients of (\ref{eq8a}).  }
\label{Table:Table3}
    \begin{tabular}{|c|c|c|c|c|c|c|}
    \hline
$\gamma'$ & $\delta'$ & $T'_{c}$ & $p'_{c}$ & $\tilde{G}_{0}$ & $\tilde{G}_{1}$ & $\tilde{G}_{2}$\\
\hline
1,114 & 0,499 & 32,186	 & 25,849 & -784,41 & 764,05	& -276,06\\
\hline\hline
$\tilde{G}_{3}$&$\tilde{G}_{4}$& $\tilde{D}_{1}$ &$\tilde{D}_{2}$ &$\tilde{D}_{3}$ &$\tilde{D}_{4}$ & -\\ 
\hline
43,412 &	-2,4315& 3,6486&2,9005 &2,8373&	3,975 & - \\
\hline
  \end{tabular}
\end{center}
\end{table} 

\begin{table}
\begin{center}
\scriptsize
\caption {Somaloy 500, high-losses phase. Values of the induction pseudo-state equation's parameters and the Pad{\'e} approximant's coefficients of (\ref{eq8a}). }
\label{Table:Table4}
    \begin{tabular}{|c|c|c|c|c|c|c|}
    \hline
$\gamma'$ & $\delta'$ & $T'_{c}$ & $p'_{c}$ & $\tilde{G}_{0}$ & $\tilde{G}_{1}$ & $\tilde{G}_{2}$\\
\hline
1,1146 & 0,4992 & 32,19 & 25,83 & -808,91 &747,44 &-266,95\\
\hline\hline
$\tilde{G}_{3}$&$\tilde{G}_{4}$& $\tilde{D}_{1}$ &$\tilde{D}_{2}$ &$\tilde{D}_{3}$ &$\tilde{D}_{4}$ & -\\ 
\hline
42,852 &	-2,5142 & 0,5846 & -3,001 &-9,968 &	7,1432 & - \\
\hline
  \end{tabular}
\end{center}
\end{table} 
\section{Optimization of Induction and Power Losses}
\begin{figure}
\begin{center}
\includegraphics[ width=10cm]{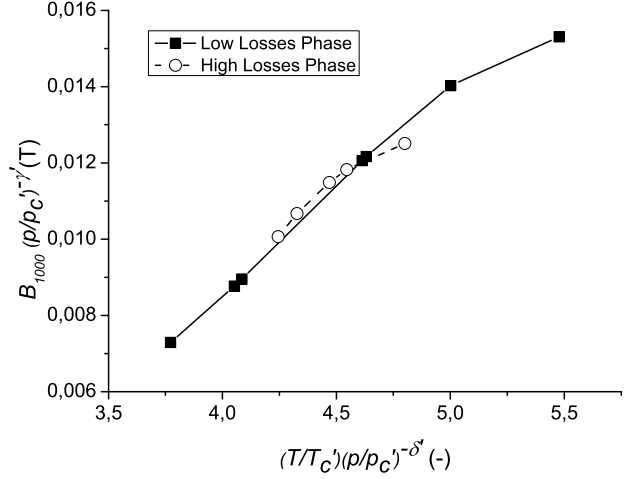}
\caption{Scaled $B_{1000}$ vs. scaled temperature in the low loss and high loss phases.}
\label{Fig.1}
\end{center}
\end{figure} 

\begin{figure}
\begin{center}
\includegraphics[ width=10cm]{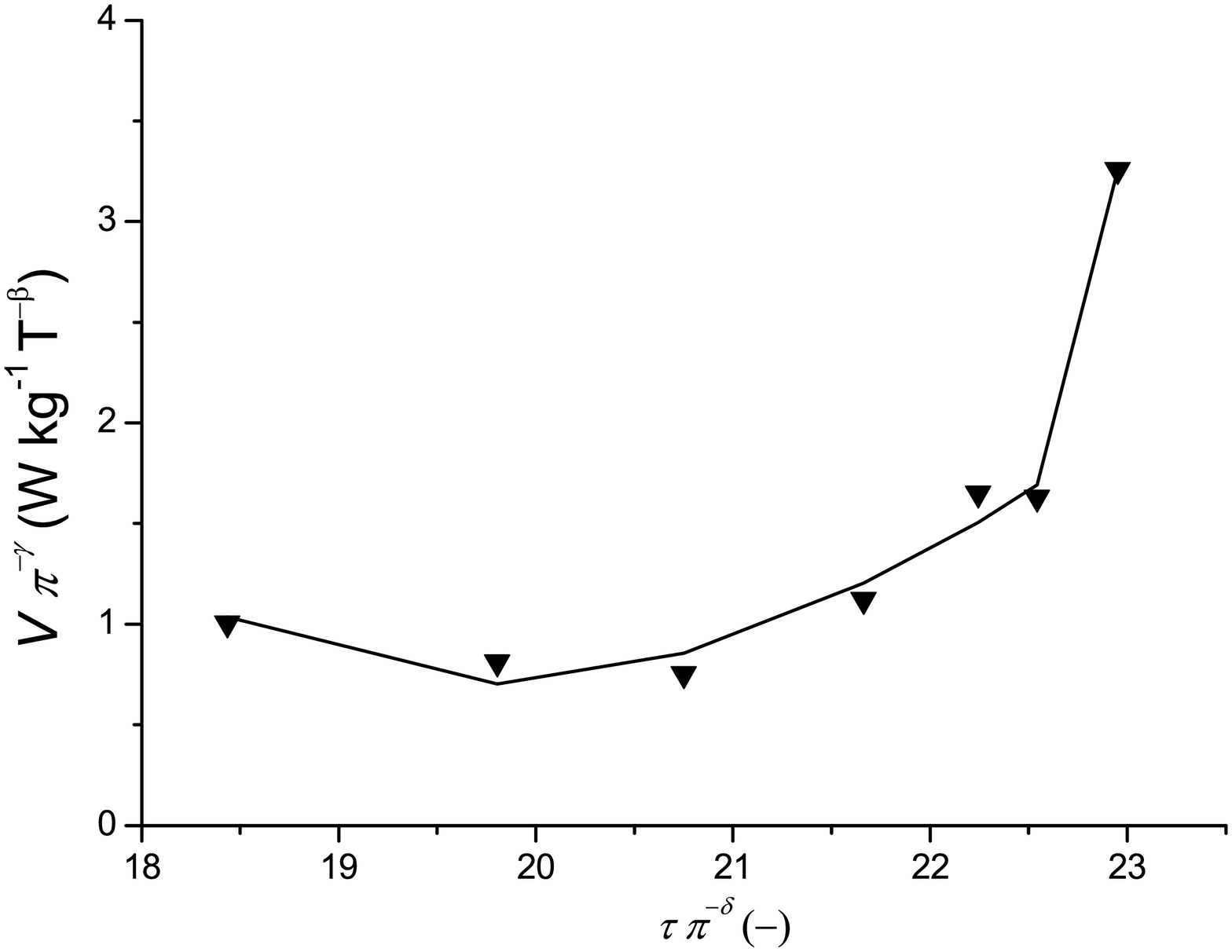}
\caption{Scaled $V$ vs. scaled temperature in the low losses phase. According to (\ref{general2a}) $\tau\,\pi^{-\delta}=X$.}
\label{Fig.2}
\end{center}
\end{figure}

In the optimization of the power loss problem \citep{bib:sok2} we have applied low loss phase solutions and high loss phase solutions have not been considered. However, it is not clear whether this simplification excludes important solutions for the induction. The binary relations are invariant with respect to scaling \citep{bib:egenh},\citep{bib:sok2}. This enables us to present all scaled characteristics in the one picture Fig.\ref{Fig.1} and draw the following conclusion. All considered pressure characteristics of the high losses phase are covered by the set of the low losses phase characteristics. Therefore for further investigations we limit our searching to the low losses phase. To this end we draw part of the phase diagram of Somaloy 500 corresponding to the low losses phase Fig. \ref{Fig.2} and we deliver values of $G_{i}, D_{i},p_{c},\gamma$ which are displayed in Table \ref{Table:TableA}.\\

\begin{table}
\begin{center}
\scriptsize
\caption {Somaloy 500, low-losses phase. Values of the $V$ pseudo-state equation's parameters and the Pad{\'e} approximant's coefficients of (\ref{eq8a}) \citep{bib:sok2}. This table is an amended version of Table 3 in \citep{bib:sok2} }

\label{Table:TableA}
    \begin{tabular}{|c|c|c|c|c|c|c|}
    \hline
$\gamma$ & $\delta$ & $T_{c}$ & $p_{c}$ & $G_{0}$ & $G_{1}$ & $G_{2}$\\
\hline
 1,2812 & 0,1715 &21,622	 & 37,729 & 370315315 & -47752251	& 1734952\\
\hline\hline
$G_{3}$&$G_{4}$& $D_{1}$ &$D_{2}$ &$D_{3}$ &$D_{4}$ & -\\ 
\hline
-1,3764 &	-678,26 & 170,80 & 6243,8 &	386,96  &	-28,699 & - \\
\hline
  \end{tabular}
\end{center}
\end{table} 

\begin{table}
\begin{center}
\scriptsize
\caption {Somaloy 500, low-losses phase.  Optimum solutions in technological and in physical spaces. }
\label{Table:TableB}
    \begin{tabular}{|c|c|c|c|}
    \hline
$p\textrm{(MPa)} $& $T(^o\textrm{C})$ &$V(\textrm{W/kgT}^{-\beta})$ & $B_{1000}\textrm{(T)}$ \\
\hline

389	&370	&14,1&	0,300\\
492	&407	&20,0	&0,356\\
584&	440	&27,3	&0,400\\
683&	478	&40,0	&0,449\\
733&	499&	50,0	&0,479\\
764	&515&	58,5&	0,500\\
800	&532	&70,0	&0,525\\
838&	549	&82,9&0,550\\
906&	570	&101&	0,580\\
979	&584&	116	&0,600\\

\hline
  \end{tabular}
\end{center}
\end{table} 

\begin{figure}
\begin{center}
\includegraphics[ width=10cm]{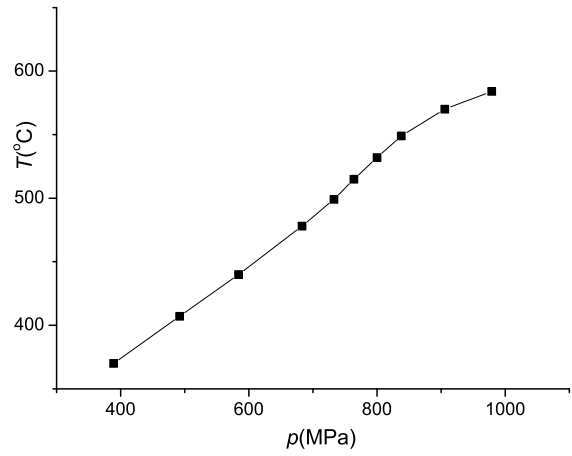}
\caption{Technological optimum curve presenting dependence of optimum temperature vs optimum pressure.}
\label{Fig.3}
\end{center}
\end{figure} 

\begin{figure}
\begin{center}
\includegraphics[ width=10cm]{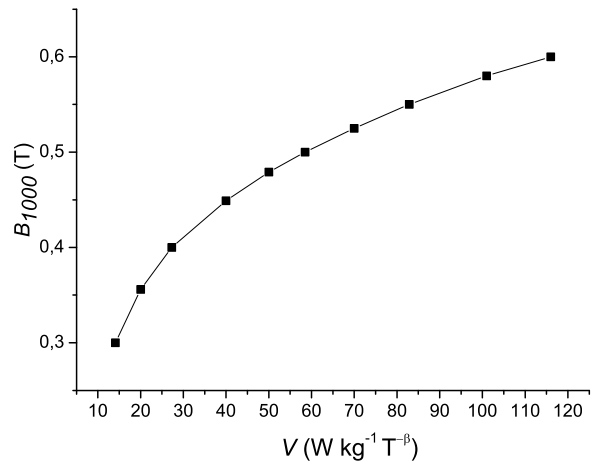}
\caption{Physical optimum curve. Induction $B_{1000}$ vs.losses measure $V$.}
\label{Fig.4}
\end{center}
\end{figure} 
\section{Details of Bicriteria Problem}
\begin{figure}
\begin{center}
\includegraphics[ width=10cm]{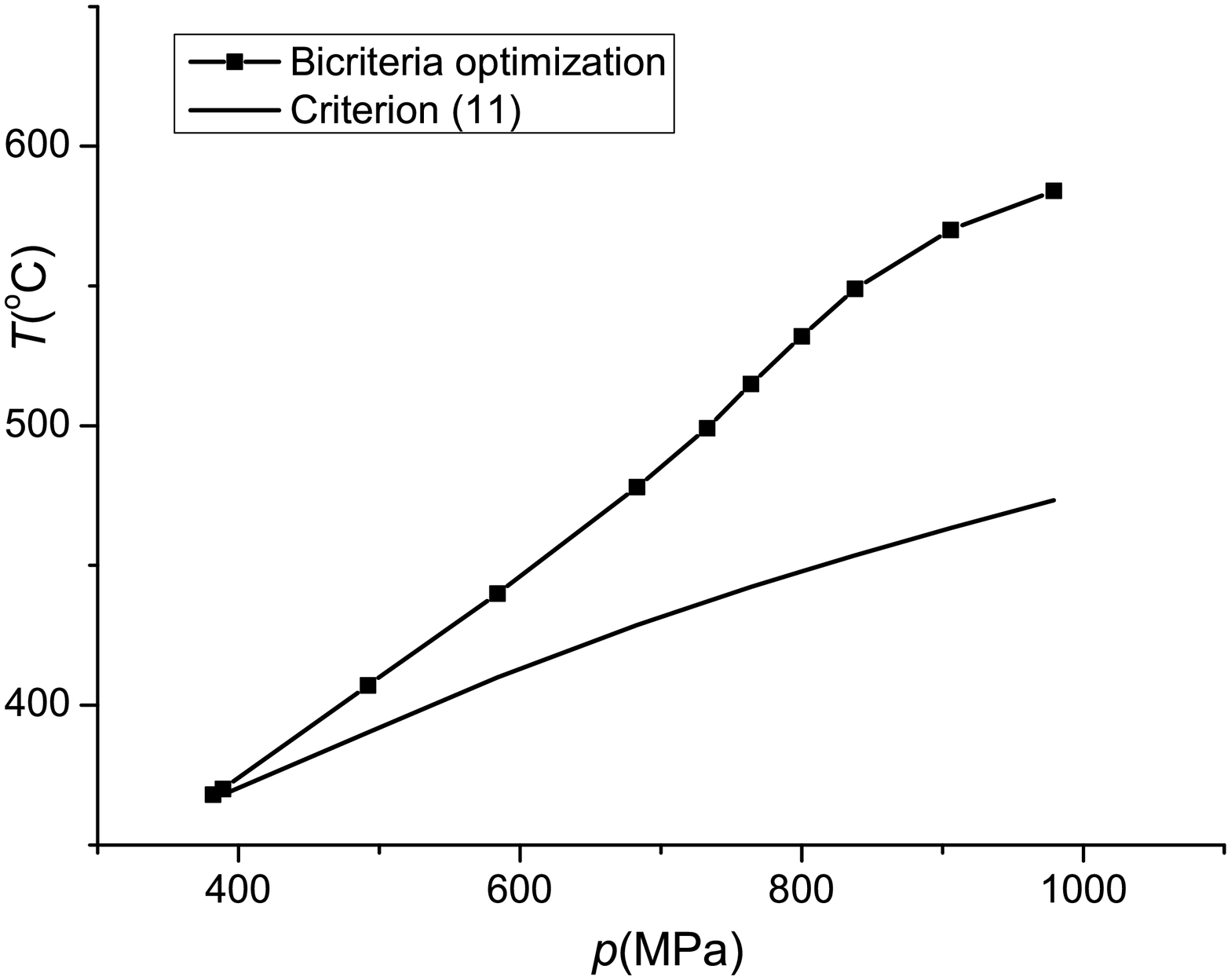}
\caption{Reduction of the feasible set of solutions for the technological parameters $\{p,T\}$ to the single point  $(p=382 \textrm{MPa}, T=368^{o}\textrm{C})$.}
\label{Fig.4NOWY}
\end{center}
\end{figure}

All calculations in this section have to satisfy  the following conditions: $18,4<X<22,9$, which results from limitation of the presented calculations to the Low Losses Phase presented in Fig. \ref{Fig.2}.
The considered bicriteria problem is formulated by the  initial value of $V=V_{1}$, feasible set of $(p,T)$  and the two criteria : $V(p,T)=V_{min}$ whereas $B_{1000}(p,T)=B_{1000\,max}$.
 Since increase of $B_{1000}$ causes increase of $V$ these conditions are in contradiction. 
 Therefore  looked for solving criterion should lead to  self- consistency between $V_{min}$ and $B_{1000\,max}$.  Such  consistency will be achieved as a fixed point of the following recurrence procedure.
Let maximiztion and minimization procedures be reprezented by the operators $\hat O_{max}$ and $\hat O_{min}$, respectively. Let $B_{1000}(T,p)$ and $V(T,p)$  be functions defined by $(\ref{eq8a})$ and $(\ref{eq8aa})$ respectively. Then the one step of  independent optimizations  of $B_{1000}(T,p)$ and $V(T,p)$ can be writen in the following form:
\begin{eqnarray}
\hat {O}_{max} B_{1000}=B_{1000,max}\hspace{2mm}for\hspace{2mm} T=T_{1},\hspace{2mm} p=p_{1} \label{R2}\\
\hat {O}_{min}V=V_{min}\hspace{2mm}for\hspace{2mm} T=T_{2},\hspace{2mm} p=p_{2} \label{R3}.
\end{eqnarray}
The obtained result consits of the two points $(T_{1},p_{1}),(T_{2},p_{2})$ and the corresponding values of magnitudes to be optimized $B_{1000}(T_{1},p_{1})$ and $V(T_{2},p_{2})$. Therefore, any further optimization is not possible and the bicriteria problem is not solved. 
In order to mesh $B_{1000,max}$ and $V_{min}$ we introduce constrain $V(T,p)=V_{0}$  which protects (\ref{R2}) and (\ref{R3}) against collapse, where $V_{0}$ is an initial value of loss. Then (\ref{R2}) get the  following  form which coupled $B_{1000,max}$ and $V_{min}$  as well as leaved some space for further optimization:
\begin{equation}
\label{R4}
V(T,p)=V_{0},\hspace{2mm}\hat {O}_{max} B_{1000}(T,p)=B_{1000}(T_{1},p_{1}).
\end{equation} 
Having $B_{1000}(T_{1},p_{1})$ we protect (\ref{R3}) against collapse:
\begin{equation}
\label{R5}
B_{1000}(T,p)=B_{1000}(T_{1},p_{1}),\hspace{2mm}\hat {O}_{min}V=V(T_{2},p_{2}).
\end{equation} 
Therefore, after $n$ steps we obtain:
\begin{eqnarray}
V(T,p)=V(T_{2\,n},p_{2\,n}),\hspace{2mm}\hat {O}_{max} B_{1000}(T,p)=B_{1000}(T_{2\,n+1},p_{2\,n+1})\label{R6},\\
B_{1000}(T,p)=B_{1000}(T_{2\,n+1},p_{2\,n+1}),\hspace{2mm}\hat {O}_{min}V=V(T_{2\,n+2},p_{2\,n+2})\label{R7}.
\end{eqnarray}
(\ref{R4})-(\ref{R7}) generate the two converging series: $T_{1},T_{2},\cdots,T_{2\,n+2}$ and $p_{1},p_{2},\cdots,p_{2\,n+2}$:
\begin{eqnarray}
\lim\limits_{k\to\infty}T_{k}=T^{*}\label{fix1}\\
\lim\limits_{k\to\infty}p_{k}=p^{*}\label{fix2}.
\end{eqnarray}
Substituting $T^{*}$ and $p^{*}$ to $(\ref{eq8a})$ and $(\ref{eq8aa})$ we derive the meshed values of $V$ and $B_{1000}$:
\begin{equation}
\label{fix3}
V^{*}=V(T^{*},p^{*}),\hspace{2mm} B_{1000}^{*}=B_{1000}(T^{*},p^{*}).
\end{equation}
The found solutions are not unique. Selecting set of initial values for $V_{o}$ we derive the set of final solutions. 
Optimization has been done by SOLVER routine of EXCEL2010 program. Obtained output 
is presented in TABLE \ref{Table:TableB}. Fig.\ref{Fig.3} and Fig.\ref{Fig.4} present these results in technological and in physical spaces, respectively. The obtained results represented by markers are fixed points of the proposed procedure. There is one to one correspondence between these points in physical and technological spaces (TABLE \ref{Table:TableB}). In order to select an unique solution one must provide an additional criterion resulting from a relation between importance of losses and induction.
For instance, assuming the deepest minimum for the scaled measure of losses $V\pi^{-\gamma}$ we apply condition given by (\ref{optim}). Intersection of  two curves presented in Fig.\ref{Fig.4NOWY}  leads to the following single solution  $p=382 (\textrm{MPa})$ and $T=363(^{o}\textrm{C})$. In the physical space this point corresponds to $V=13,64(\textrm{W}\,\textrm{kg}^{-1}\textrm{T}^{-\beta})$  and $B_{1000}=0,29(\textrm{T})$.

At the end we pay some attention to the power losses measure $V$. This is an auxiliary magnitude which help us to derive values of designing technological parameters due to the following features:
\begin{itemize}
\item $V$ is pseudo-thermodynamic average with respect to magnitude created with the peak of induction and  the frequency of electromagnetic field wave. Therefore, this includes information about both independent variables.
\item $V$ depends on the technological parameters.
\item Physical dimension of $V$ is unknown yet due to a dummy exponent $\beta$.  However, the value of $V$  is well determined together with values of $p$ and $T$ which enables us to compose SMC specimen and to perform measurements of its characteristics. Finally, applying (\ref{uno}),(\ref{equ}) we are able to calculate $\beta$ and to determine the physical dimension of the current $V$.
 \end{itemize}

\section{Conclusions}
We have presented  method for the bicriteria optimization of the chosen physical properties of Soft Magnetic Composites. By this way we have solved the problem mentioned in \citep{bib:sok2} which concerns optimization of losses and induction. Achievement of the fixed point is interpreted as revelation of an equilibrium between the both assumed criteria.The crucial roles in the presented method play scaling and the notion of  pseudo-state equation. The created system is as good as the experimental data which have been used for the estimations of model parameters. Therefore,  presented here the first version will be improved by forthcoming new experimental data.  The  presented example in this paper is a minimum nontrivial case of Multiphysics problem and shows that this approach suits for designing  Magnetic Composites. Therefore,  the presented algorithm is going to be extended for  more than two physical features of the composing material.  For instance, the designing of  magnetic composites requires also optimization of mechanical properties, since the susceptibility of such materials to cracking in service is of fundamental concern \citep{bib:crack}. We address the derived algorithm  to designers of SMCs.

 \end{document}